\documentclass{PoS}

\def\mpi{M_\pi}

\title{Nuclear forces on the lattice}

\ShortTitle{Nuclear forces on the lattice}

\author{\speaker{Silas Beane}%
\\
        University of New Hampshire \\
        E-mail: \email{silas@physics.unh.edu}}

\abstract{Recent studies by the NPLQCD collaboration of hadronic
  interactions using lattice QCD are reviewed, with an emphasis on a
  recent calculation of meson-baryon scattering lengths. Ongoing
  high-statistics calculations of baryon interactions are also
  reviewed. In particular, new insights into the signal/noise problems
  that plague correlation functions involving baryons are discussed.}

\FullConference{6th International Workshop on Chiral Dynamics\\
                 July 6-10 2009\\
                 Bern, Switzerland}

\begin{document}

\section{Introduction}

\noindent Quantum Chromodynamics (QCD) ---the underlying theory of the
strong interactions--- when combined with the electroweak
interactions, is responsible for all of nuclear physics, from the
structure of light nuclei to the most complex nuclear
reactions. Presently lattice QCD is the only known method which allows
one to compute properties of hadrons and nuclei from first
principles. While lattice QCD provides highly-accurate predictions and
postdictions of hadronic properties like masses and decay constants,
the study of hadronic interactions, like two-body scattering and
three-body interactions is still in a development phase. The NPLQCD
collaboration has calculated the interactions between many of the
lowest-lying baryons and mesons, including $\pi^+\pi^+$, $K^+\pi^+$,
$K^+K^+$, meson-baryon, nucleon-nucleon, and hyperon-nucleon
scattering as well as multi-meson systems (much of this work is
reviewed in Ref.~\cite{Beane:2008dv}). It is noteworthy that the
$\pi^+\pi^+$ scattering length has been calculated to percent-level
accuracy at the physical point~\cite{Beane:2007xs}. Part of the reason
behind this accuracy is that QCD correlation functions involving
Goldstone bosons are subject to powerful chiral symmetry
constraints. Since current lattice calculations are carried out at
unphysical quark masses, these constraints play an essential role in
extrapolating the lattice data to the physical quark masses, as well
as to the infinite volume, and continuum limits. Chiral perturbation
theory ($\chi$-PT) is the optimal method for implementing QCD
constraints due to chiral symmetry, and in essence, provides an
expansion of low-energy S-matrix elements in quark masses and powers
of momentum~\cite{Bernard:2007zu}.

In contrast to the purely mesonic sector, recent studies of
baryon-baryon interactions ---the paradigmatic nuclear physics
process--- have demonstrated the fundamental difficulty faced in
making predictions for baryons and their
interactions~\cite{Beane:2006mx,Beane:2006gf}. Unlike the case with
mesons, correlation functions involving baryons suffer an exponential
degradation of signal/noise at large times~\cite{Lepage:1989hd}.
Furthermore, while baryon interactions are constrained by QCD
symmetries like chiral symmetry, the constraints are not nearly as
powerful as when there is at least one pion or kaon in the initial or
final state. For instance, there is no expectation that the
baryon-baryon scattering lengths vanish in the chiral limit as they do
in the purely mesonic sector. To make matters worse, in s-wave
nucleon-nucleon scattering, the interactions are enhanced due to the
close proximity of a non-trivial fixed point of the renormalization
group, which drives the scattering lengths to infinity, thus rendering
the effective field theory description of the interaction
non-perturbative~\cite{Bedaque:2002mn}.

In this proceedings I will review the most recent progress made in the
description of hadronic scattering and interactions.  Given the
contrast in difficulty between the purely mesonic and purely baryonic
sectors described above, it is clearly of great interest to perform a
lattice QCD investigation of the simplest scattering process involving
at least one baryon: meson-baryon scattering. The first
fully-dynamical study of meson-baryon scattering has been carried out
recently in Ref.~\cite{Torok:2009dg} and will be reviewed.  I will
also review other recent work~\cite{Beane:2009ky,Beane:2009gs} by the
NPLQCD collaboration, which has performed very-high statistics
calculations of few-body baryon systems that offer new insights into
the signal/noise problem, and allow a first glimpse of the baryon
number $B=3$ sector.

\section{Signal to Noise Estimates}

\noindent As is well known~\cite{Lepage:1989hd}, very general
field-theoretic arguments allow a robust estimate of the noise to
signal ratio of hadronic correlation functions calculated on the
lattice. With an eye towards lattice QCD attempts to describe nuclei,
it is worth briefly noting the fundamental difference between
lattice-measured correlation functions involving mesons and
baryons. As an example, consider the noise to signal ratio of a
correlation function involving $n$ pion fields, where the small
interaction is neglected,
\begin{eqnarray}
{\sigma (t)\over \langle \theta(t) \rangle }&&
\sim \ {\sqrt{\left(A_2 - A_0^2 \right)} \ e^{-n m_\pi t}\over \sqrt{N} A_0\
e^{-n m_\pi t} }
\sim {1\over \sqrt{N}}
 \ \ .
\label{eq:NtoSpi}
\end{eqnarray}
Here $\langle \theta(t)\rangle$ is the correlation function, $\sigma
(t)$ is the variance and the $A_i$ are amplitudes. It is noteworthy
that in this ratio, the time dependence of the variance mirrors the
time dependence of the correlator itself. One therefore concludes that
correlators involving arbitrary numbers of pions have time-independent
errors, as is indeed observed in lattice calculations. This of course
renders the study of mesonic correlators quite pleasurable (from the 
statistical perspective).

The baryons provide a more disturbing story; consider the noise to signal
ratio for a proton correlation function:
\begin{eqnarray}
{\sigma (t)\over \langle \theta(t) \rangle }&&
\sim \ {\sqrt{A_2} \ e^{-{3\over 2} m_\pi t}\over \sqrt{N} A_0 e^{- m_p t} }
\sim   {1\over \sqrt{N}} \ e^{\left( m_p - {3\over 2} m_\pi\right) t} 
 \ \ .
\label{eq:NtoSproton}
\end{eqnarray}
Here the variance is dominated by the three-pion state rather than by
the proton, and therefore the noise to signal ratio of the proton
correlator grows exponentially with time.  More generally, for a
system of $A$ nucleons, the noise to signal ratio behaves as
\begin{eqnarray}
{\sigma (t)\over \langle \theta(t) \rangle }&&
\sim   {1\over \sqrt{N}} \ e^{\ A \left( m_p - {3\over 2} m_\pi\right) t}
 \ \ .
\label{eq:NtoSnucleus}
\end{eqnarray}
Therefore the situation worsens as one adds nucleons. Indeed one can
show that the theoretical expectation from eq.~\ref{eq:NtoSnucleus}
compares favorably with $A=2$ data calculated by the NPLQCD
collaboration~\cite{Bedaque:2007pe}.

The estimates given above, which follow from very general field
theoretic arguments, naively indicate that nucleon and nuclear physics
require {\it exponentially more resources} than meson physics to
achieve the same level of accuracy. While these general results are
somewhat discouraging to efforts to extract baryon interactions from
the lattice, one should keep in mind that the arguments are rigorous
only in the asymptotic, infinite time, limit. And, indeed, as we shall
see below, there can be a region of intermediate times where these
arguments do not apply, rendering the problem much less severe than
these arguments suggest.

\section{Meson-meson interactions}

\begin{figure}[hb!]
\begin{center}
\centerline{\includegraphics*[width=0.7\textwidth]{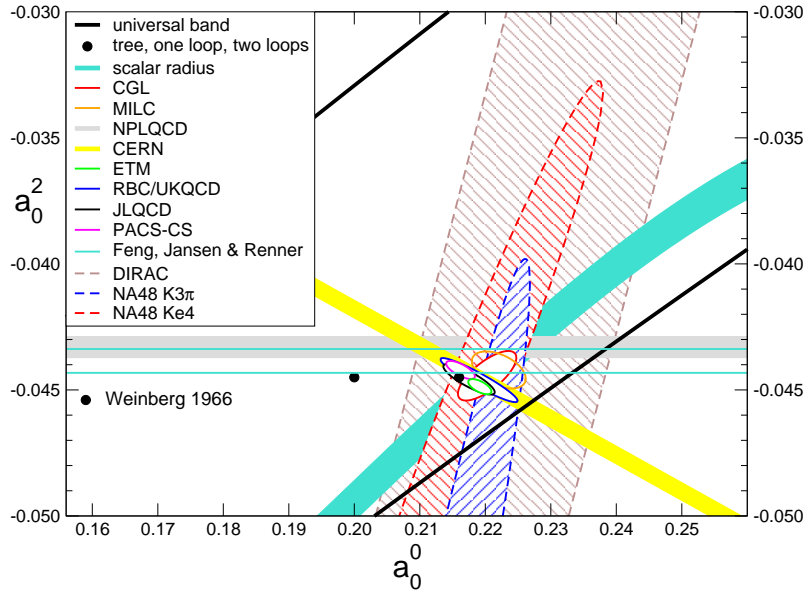}}
\caption{The state of threshold s-wave $\pi\pi$ scattering. Noteworthy are
the red ellipse from the Roy equation analysis and the grey band from the direct lattice QCD calculation of
the $I=2$ scattering length, as discussed in the text.} \label{fig:pipi2}
\end{center}
\end{figure}

\noindent Here I will briefly review a calculation of the $I=2$
pion-pion ($\pi\pi$) scattering length, as it serves as a benchmark
calculation with an accuracy that can only be aspired to at present for other systems.
Due to the chiral symmetry of QCD, pion-pion $\pi\pi$ scattering at
low energies is the simplest and best-understood hadron-hadron
scattering process.  The scattering lengths for $\pi\pi$ scattering in
the s-wave are uniquely predicted at leading order in chiral
perturbation theory ($\chi$-PT)~\cite{Weinberg:1966kf}:
\begin{eqnarray}
m_\pi a_{\pi\pi}^{I=0} \ = \ 0.1588 \ \ ; \ \ m_\pi a_{\pi\pi}^{I=2} \ = \
-0.04537 
\ \ \ ,
\label{eq:CA}
\end{eqnarray}
at the charged pion mass. While experiments do not provide stringent
constraints on the scattering lengths, a determination of s-wave
$\pi\pi$ scattering lengths using the Roy equations has reached a
remarkable level of
precision~\cite{Colangelo:2001df,Leutwyler:2008fi}:
\begin{eqnarray}
m_\pi a_{\pi\pi}^{I=0} \ = \ 0.220\pm 0.005 \ \ ; \ \ m_\pi a_{\pi\pi}^{I=2} \ = \ -0.0444\pm 0.0010
\ \ \ .
\label{eq:roy}
\end{eqnarray}
The Roy equations~\cite{Roy:1971tc} use dispersion theory to relate
scattering data at high energies to the scattering amplitude near
threshold. At present, lattice QCD can compute $\pi\pi$ scattering
only in the $I=2$ channel as the $I=0$ channel contains disconnected
diagrams. It is of course of great interest to compare the precise Roy
equation predictions with lattice QCD
calculations. Figure~\ref{fig:pipi2} summarizes theoretical and
experimental constraints on the s-wave $\pi\pi$ scattering
lengths~\cite{Leutwyler:2008fi}. It is clearly a strong-interaction
process where theory has outpaced the very-challenging experimental
measurements.

The only existing fully-dynamical lattice QCD prediction of the $I=2$
$\pi\pi$ scattering length involves a mixed-action lattice QCD scheme
of domain-wall valence quarks on a rooted staggered sea. Details of
the lattice calculation can be found in Ref.~\cite{Beane:2007xs}.
The scattering length was computed at pion masses, $m_\pi\sim 290~{\rm
  MeV}$, $350~{\rm MeV}$, $490~{\rm MeV}$ and $590~{\rm MeV}$, and at
a single lattice spacing, $b\sim 0.125~{\rm fm}$ and lattice size
$L\sim 2.5~{\rm fm}$~\cite{Beane:2007xs}. The physical value of the
scattering length was obtained using two-flavor mixed-action $\chi$-PT
which includes the effect of finite lattice-spacing
artifacts to $\mathcal{O}(m_\pi^2 b^2)$ and
$\mathcal{O}(b^4)$~\cite{Chen:2006wf}.  The final result is:
\begin{eqnarray}
m_\pi a_{\pi\pi}^{I=2} & = &  -0.04330 \pm 0.00042
\ \ \ ,
\label{eq:nplqcd2}
\end{eqnarray}
where the statistical and systematic uncertainties have been combined
in quadrature.  Notice that $1\%$ precision is claimed in this
result. The agreement between this result and the Roy equation
determination is a striking confirmation of the lattice methodology,
and a powerful demonstration of the constraining power of chiral
symmetry in the meson sector. It would clearly be of great interest to
see other (fully-dynamical) lattice QCD calculations of the s-wave
$\pi\pi$ scattering lengths using different types of fermions.  The
$K^+K^+$ and $\pi^+K^+$ scattering lengths have also been computed by
the NPLQCD collaboration. I refer the interested reader to
Ref.~\cite{Beane:2008dv} for details.


\section{Meson-Baryon Interactions}

\noindent Pion-nucleon scattering has long been considered a
paradigmatic process for the comparison of $\chi$-PT and experiment.
To this day, controversy surrounds determinations of the pion-nucleon
coupling constant and the pion-nucleon sigma term. The $K^- n$
interaction is important for the description of kaon condensation in
the interior of neutron stars~\cite{KaplanNelson}, and meson-baryon
interactions are essential input in determining the final-state
interactions of various decays that are interesting for standard-model
phenomenology (See Ref.~\cite{Lu:1994ex} for an example). In
determining baryon excited states on the lattice, it is clear that the
energy levels that represent meson-baryon scattering on the
finite-volume lattice must be resolved before progress can be made
regarding the extraction of single-particle excitations.  

While pion-nucleon scattering is the best-studied meson-baryon
process, both theoretically and experimentally, its determination on
the lattice is computationally prohibitive since it involves
annihilation diagrams. At present only a few limiting cases that
involve these diagrams are being investigated~\cite{Babich:2009rq}.
Combining the lowest-lying $SU(3)$ meson and baryon octets, one can
form five meson-baryon elastic scattering processes that do not
involve annihilation diagrams: $\pi^+\Sigma^+$, $\pi^+\Xi^0$, $K^+ p$,
$K^+ n$, and ${\bar K}^0 \Xi^0$. (Note that ${\bar K}^0 \Sigma^+$ has
the same quantum numbers as $\pi^+\Xi^0$.)  Three of these processes
involve kaons and therefore are, in principle, amenable to an $SU(3)$
heavy-baryon $\chi$-PT (HB$\chi$-PT) analysis~\cite{Jenkins:1990jv}
for extrapolation. The remaining two processes involve pions
interacting with hyperons and therefore can be analyzed in conjunction
with the kaon processes in $SU(3)$ HB$\chi$-PT, or independently using
$SU(2)$ HB$\chi$-PT. Below we will review recent work in
Ref.~\cite{Torok:2009dg} which has carried out this analysis.

The calculations of Ref.~\cite{Torok:2009dg} were performed
predominantly with the coarse MILC lattices with a lattice spacing of
$b\sim 0.125$~fm, and a spatial extent of $L\sim 2.5$~fm.  On these
configurations, the strange quark was held fixed near its physical
value while the degenerate light quarks were varied over a range of
masses. (These are the same resources that were used to calculate
the $\pi^+\pi^+$ scattering length described above.)

\begin{figure}[hb!]
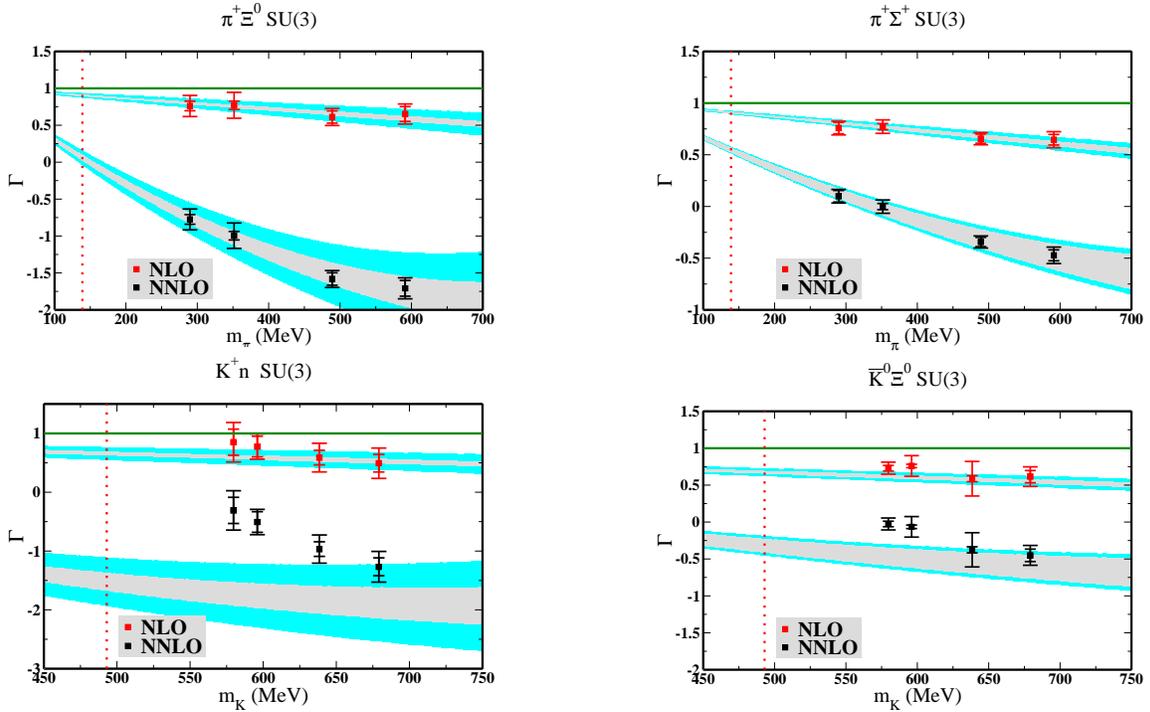

\begin{center}
\centerline{\includegraphics[width=6.5 cm]{Pi_Xi_SU3.eps}\hfill\includegraphics[width=6.5 cm]{Pi_Sigma_SU3.eps}}
\centerline{\includegraphics[width=6.5 cm]{Kaon_Neutron_SU3.eps}\hfill\includegraphics[width=6.5 cm]{Kaon_Xi_SU3.eps}}
\caption{Plots of $\Gamma_{NLO}$ and $\Gamma_{NNLO}$ versus the
  Goldstone masses for four meson-baryon processes in the SU(3) case. The line at $\Gamma=1$ is the leading
  order curve, and dotted line is the physical meson mass. }
\label{fig:KostasNLONNLO}
\end{center}
\end{figure}

The scattering lengths of the five meson-baryon processes without
annihilation have been calculated to $\mathcal{O}(m_{\pi,K}^3)$ in $SU(3)$
HB$\chi$-PT~\cite{Liu:2006xja,Liu:2007ct}, and are given by
\begin{eqnarray}
a_{\pi^+\Sigma^+}=\frac{1}{4 \pi}\frac{m_\Sigma}{m_\pi+m_\Sigma} \bigg[ -\frac{2m_\pi}{f_\pi^2} + \frac{2m_\pi^2}{f_\pi^2}C_1 + \mathcal{Y}_{\pi^+\Sigma^+}(\mu ) 
+ 8 h_{123}(\mu )\frac{m_\pi^3}{f_\pi^2} \bigg] \ ;
\label{eq:apisigfull}
\end{eqnarray}
\begin{eqnarray}
a_{\pi^+\Xi^0}=\frac{1}{4 \pi}\frac{m_\Xi}{m_\pi+m_\Xi} \bigg[ -\frac{m_\pi}{f_\pi^2} + \frac{m_\pi^2}{f_\pi^2}C_{01} + \mathcal{Y}_{\pi^+\Xi^0}(\mu ) + 8 h_1(\mu )\frac{m_\pi^3}{f_\pi^2} \bigg] \ ;
\label{eq:apixifull}
\end{eqnarray}
\begin{eqnarray}
a_{K^+ p}=\frac{1}{4 \pi}\frac{m_N}{m_K+m_N} \bigg[ -\frac{2m_K}{f_K^2} + \frac{2m_K^2}{f_K^2}C_1 + \mathcal{Y}_{K^+ p}(\mu ) + 8 h_{123}(\mu )\frac{m_K^3}{f_K^2} \bigg] \ ;
\label{eq:akpfull}
\end{eqnarray}
\begin{eqnarray}
a_{K^+ n}=\frac{1}{4 \pi}\frac{m_N}{m_K+m_N} \bigg[ -\frac{m_K}{f_K^2} + \frac{m_K^2}{f_K^2}C_{01} + \mathcal{Y}_{K^+ n}(\mu ) + 8 h_1(\mu )\frac{m_K^3}{f_K^2} \bigg] \ ;
\label{eq:aknfull}
\end{eqnarray}
\begin{eqnarray}
a_{\overline{K}{}^0 \Xi^0}=\frac{1}{4 \pi}\frac{m_\Xi}{m_K+m_\Xi} \bigg[ -\frac{2m_K}{f_K^2} + \frac{2m_K^2}{f_K^2}C_1 + \mathcal{Y}_{\overline{K}{}^0 \Xi^0}(\mu ) 
+ 8 h_{123}(\mu )\frac{m_K^3}{f_K^2} \bigg] \ ,
\label{eq:akxifull}
\end{eqnarray}
where the $C$'s and the $h$'s are low-energy constants (LECs) and the $\mathcal{Y}$'s are loop functions given in
Ref.~\cite{Torok:2009dg}. It is clear that this is an overconstrained system. Many fitting strategies
are possible, as discussed in Ref.~\cite{Torok:2009dg}.  We found it convenient to rewrite the
chiral perturbation theory formulas as polynomial expansions. For instance, 
in the case of $\pi^+\Xi^0$ and $K^+n$, we formed the objects:
\begin{eqnarray}
\Gamma_{NLO}&&\equiv-\frac{4 \pi a f_\phi^2}{m_\phi}\bigg(1 + \frac{m_\phi}{m_B}\bigg)=1-{ C_{01}} m_\phi \\
\Gamma_{NNLO}&&\equiv
-\frac{4 \pi a f_\phi^2}{m_\phi}\bigg(1 + \frac{m_\phi}{m_B}\bigg) + \frac{f_\phi^2}{m_\phi}\mathcal{Y}_{\phi B}({ \Lambda_{\chi}})
=1-{ C_{01}} m_\phi-8{ h_1(\Lambda_{\chi})} m_\phi^2 
\label{eq:gammadef}
\end{eqnarray}
for each of the processes. Notice that the left-hand sides of these equations are given entirely
in terms of lattice-determined quantities, while the right-hand side provides a convenient
polynomial fitting function. Figure~\ref{fig:KostasNLONNLO} plots these
functions for several of the processes.  The shift of the value of
$\Gamma$ from NLO to NNLO is dependent on the renormalization scale
$\mu$, and therefore with the choice $\mu=\Lambda_\chi$ one would
expect this shift to be perturbative if the expansion is
converging. The large shifts in $\Gamma$ from NLO to NNLO are
indicative of large loop corrections.
The LECs fit to the lattice data are tabulated in
Ref.~\cite{Torok:2009dg}.  While the NNLO LECs $h_1$ and $h_{123}$
appear to be of natural size, the NLO LECs $C_0$ and $C_{01}$ are
unnaturally large. The extrapolated values of the five scattering
lengths are given in Table~\ref{tab:scattLpisig}. While the
$\pi^+\Sigma^+$ and $\pi^+\Xi^0$ scattering lengths appear to be
perturbative, the extrapolated kaon-baryon scattering lengths at NNLO
deviate by at least 100\% from the LO values.  The seemingly
inescapable conclusion is that the kaon-baryon scattering lengths are
unstable against chiral corrections in the three-flavor chiral
expansion, over the range of light-quark masses that we consider.
\begin{table}
\begin{center}
\begin{tabular}{ccccc}
Quantity & LO (fm) & NLO fit (fm) & NLO (NNLO fit) (fm) & NNLO (fm) \\
\hline
$a_{\pi\Sigma}$ & -0.2294 & -0.208(01)(03) & -0.117(06)(08) & -0.197(06)(08) \\
$a_{\pi\Xi}$ & -0.1158 & -0.105(01)(04) &  0.004(05)(11) & -0.096(05)(12) \\
$a_{Kp}$ & -0.3971 & -0.311(18)(44) &  0.292(35)(48) & -0.154(51)(63) \\
$a_{Kn}$ & -0.1986 & -0.143(10)(27) &  0.531(28)(68) &  0.128(42)(87) \\
$a_{K\Xi}$ & -0.4406 & -0.331(12)(31) &  0.324(39)(54) & -0.127(57)(70) \\
\end{tabular}
\end{center}
\caption{$SU(3)$ extrapolated scattering lengths. The first uncertainty in parentheses is
  statistical, and the second is the statistical and systematic uncertainty
  added in quadrature. }
\label{tab:scattLpisig}
\end{table}

\begin{figure}[!ht]
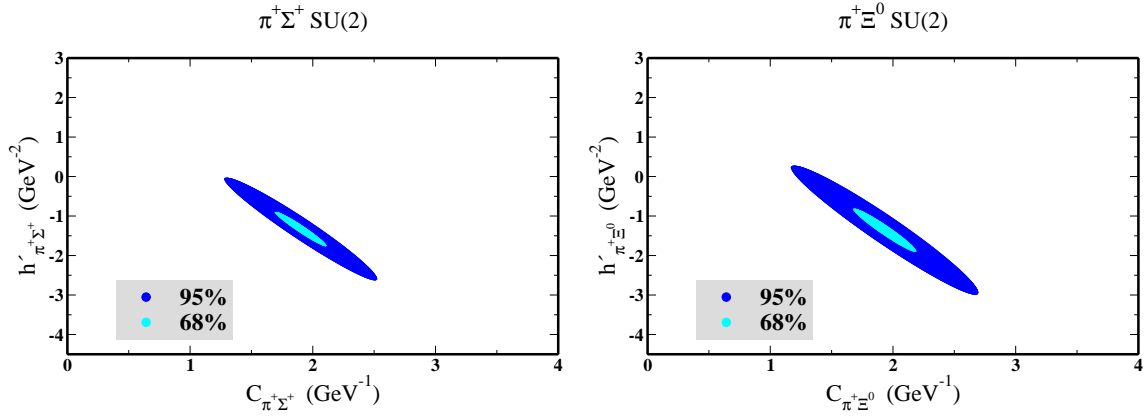

\centering
\includegraphics[width=0.49\linewidth]{ellX_Pi_Sigma_SU2.eps}\hfill
\includegraphics[width=0.49\linewidth]{ellX_Pi_Xi_SU2.eps}
\caption{The 68\% (light) and 95\% (dark) confidence interval error 
ellipses for fits for the $\pi^+\Sigma^+$ (left), and~$\pi^+\Xi^0$ (right) processes.}
\label{fig:errellipseU} 
\end{figure}

\begin{figure}[ht]
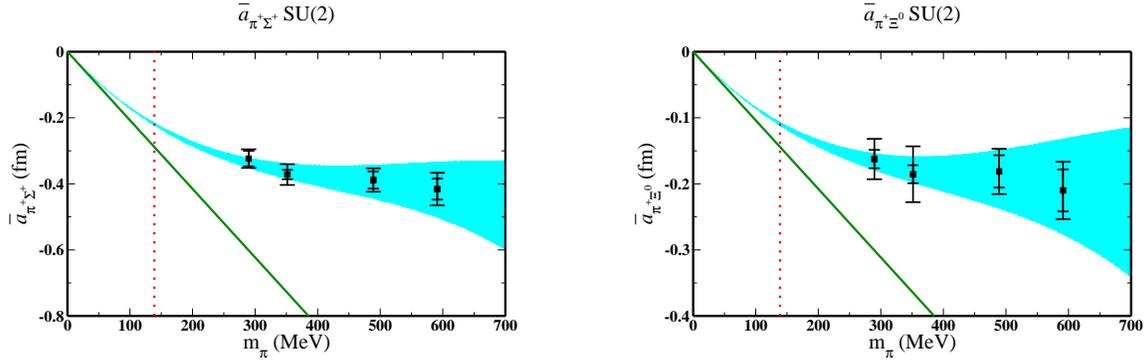

\includegraphics[width=0.45\linewidth]{a_Pi_Sigma_SU2.eps}\hfill
\includegraphics[width=0.45\linewidth]{a_Pi_Xi_SU2.eps}
\caption{$\overline{a}$ plots for the $\pi^+\Sigma^+$,
  and~$\pi^+\Xi^0$ processes versus the pion mass. The
  diagonal line is the leading order curve, and the dotted line is the
  physical pion mass. The innermost error bar is the statistical uncertainty
  and the outermost error bar is the statistical and systematic uncertainty
  added in quadrature. The filled bands are the fits to the LECs in the
  SU(2) case at NNLO.}
\label{fig:aSU2}
\end{figure}

Given the poor convergence seen in the three-flavor chiral expansion
due to the large loop corrections, it is natural to consider the
two-flavor theory with the strange quark integrated out. In this way,
$\pi\Sigma$ and $\pi\Xi$ may be analyzed in an expansion in
$m_\pi$. To $\mathcal{O}(m_\pi^3)$ in the two-flavor chiral expansion,
one has~\cite{Mai:2009ce}
\begin{eqnarray}
a_{\pi^+\Sigma^+}=\frac{1}{2 \pi}\frac{m_\Sigma}{m_\pi+m_\Sigma} \bigg[ -\frac{m_\pi}{f^2} + \frac{m_\pi^2}{f^2} {\mathrm{C}}_{\pi^+\Sigma^+} 
+ \frac{m_\pi^3}{f^2} h'_{\pi^+\Sigma^+} \bigg], \qquad
h'_{\pi^+\Sigma^+}=\frac{4}{f^2}\ell_4^r+ h_{\pi^+\Sigma^+} \ ;
\label{eq:apisig2param}
\end{eqnarray}
\begin{eqnarray}
a_{\pi^+\Xi^0}=\frac{1}{4 \pi}\frac{m_\Xi}{m_\pi+m_\Xi} \bigg[ -\frac{m_\pi}{f^2} + \frac{m_\pi^2}{f^2}{\mathrm{C}}_{\pi^+\Xi^0} 
+ \frac{m_\pi^3}{f^2} h'_{\pi^+\Xi^0} \bigg],\qquad
h'_{\pi^+\Xi^0}=\frac{4}{f^2}\ell_4^r + h_{\pi^+\Xi^0} \ ,
\label{eq:apixi2param}
\end{eqnarray}
where $\ell_4^r$ is the LEC which governs the pion mass
dependence of $f_\pi$~\cite{Colangelo:2001df}.  Note that the chiral
logs have canceled, and in this form, valid to order
$m_\pi^3$ in the chiral expansion, the scattering lengths have a
simple polynomial dependence on $m_\pi$~\cite{Mai:2009ce}.
Figure~\ref{fig:errellipseU} shows the 68\% and 95\% confidence
interval error ellipses in the $h$-${\mathrm{C}}$ plane for both
${\pi^+\Sigma^+}$ and ${\pi^+\Xi^0}$.  Exploring the full 95\%
confidence interval error ellipse in the $h$-${\mathrm{C}}$ plane
yields
\begin{eqnarray}
a_{\pi^+\Sigma^+}&=& -0.197 \pm 0.017~{\rm fm} \ ;\\
a_{\pi^+\Xi^0}&=& -0.098\pm 0.017~{\rm fm} \ .
\label{eq:MP2}
\end{eqnarray}
These are the numbers that we quote as our best determinations of the pion-hyperon
scattering lengths.
In Figure~\ref{fig:aSU2} we plot the scattering length versus the pion mass.
(Note that the bar denotes the scattering length rescaled by a kinematical factor~\cite{Torok:2009dg}.)

\section{Baryon-Baryon Interactions}

\noindent The NPLQCD collaboration has performed the first full QCD
calculations of nucleon-nucleon interactions~\cite{Beane:2006mx} and
hyperon-nucleon~\cite{Beane:2006gf} interactions at low-energies with
large pion masses. The nucleon-nucleon scattering lengths were found
to be of natural size for the unphysically large values of the quark
masses at which these calculations were performed. In contrast, the
physical values of the scattering lengths are unnaturally large
compared to the range of the nucleon-nucleon interaction.  This
suggests a fine-tuning of the underlying parameters of QCD (the quark
masses and QCD scale) and lattice calculations with quark masses much
closer to the physical values are needed to reproduce the experimental
values. By contrast, very little is known about the interactions
between nucleons and hyperons from experiment, and lattice QCD
calculations can provide the best determinations of the corresponding
scattering parameters and hence determine the role of hyperons in
neutron stars. 

\begin{figure}[!ht]
\centering
\includegraphics[width=0.7\linewidth]{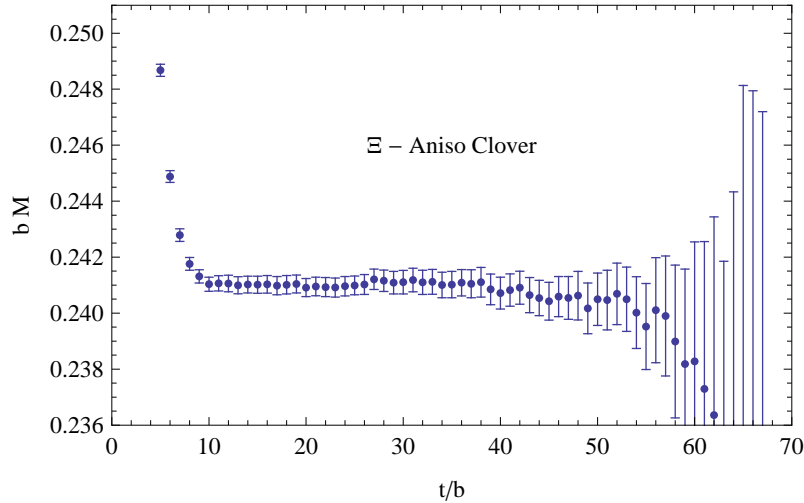}
\caption{Effective mass plot for the $\Xi$ baryon from a high-statistics
calculation using anisotropic clover gauge configurations.}
\label{fig:xiex} 
\end{figure}

The NPLQCD collaboration has undertaken an extensive exploration of
the impact of high-statistics on one-, two- and three-baryon correlation
functions on one ensemble of anisotropic clover gauge configurations
that are presently being generated by the Hadron Spectrum
Collaboration~\cite{Beane:2009ky,Beane:2009gs}.  A total of
$\sim$300,000 sets of measurements were made using $1200$ gauge
configurations of size $20^3\times 128$ with an anisotropy parameter
$\xi= b_s/b_t = 3.5$, a spatial lattice spacing of $b_s=0.1227\pm
0.0008~{\rm fm}$, and pion mass of $\mpi\sim 390~{\rm MeV}$.  The
ground state baryon masses (in lattice units) were extracted with
uncertainties that are at or below the $\sim 0.2\%$-level.  An example
of the precision that has been obtained for single baryon masses with
this calculation is shown in Figure~\ref{fig:xiex}, where the
generalized effective mass plot of the $\Xi$ baryon is shown.

\begin{figure}[hb!]
\begin{center}
\centerline{\includegraphics*[width=0.7\textwidth]{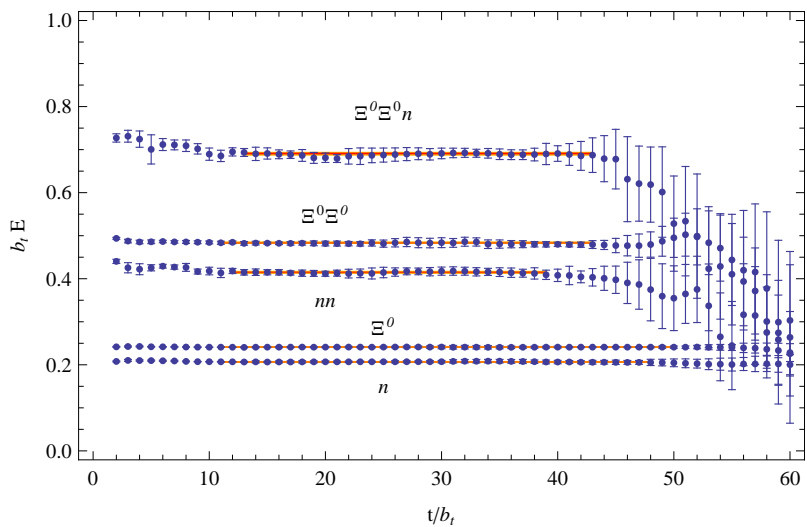}}
\caption{Effective mass plots for various one-, two- and three-body correlation functions
calculated using anisotropic clover gauge configurations. There is an intermediate region
of time slices for which signal/noise is time independent. }
\label{fig:goldenw}
\end{center}
\end{figure}

\begin{figure}[hb!]
\begin{center}
\centerline{\includegraphics*[width=0.7\textwidth]{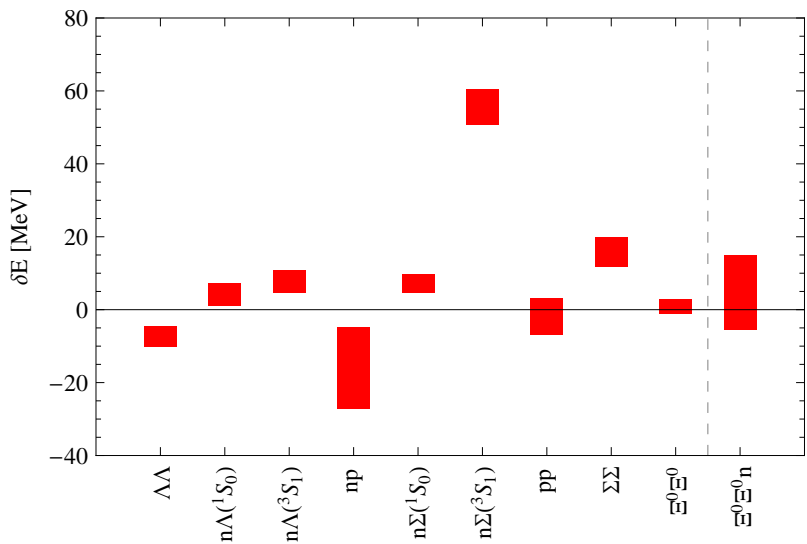}}
\caption{Energy splittings for various two- and three-body processes
calculated using anisotropic clover gauge configurations.} \label{fig:Ediff}
\end{center}
\end{figure}

With such statistics we were able to systematically explore the
signal-to-noise issue in baryon systems, and found that, while
signal/noise does degrades exponentially at large times, there is an
intermediate time-interval for which it is time-independent, producing
a noise-to-signal ratio per baryon that is essentially independent of
baryon number. This is seen clearly in Figure~\ref{fig:goldenw}.
This feature greatly increases the number of nucleons that can be
explored with lattice QCD for fixed computational resources. This was
a significant finding and opens up the possibility of calculations
with four and more baryons. In work that is currently in progress, we have
obtained a number of phase-shifts in the two-baryon sector with $>5$ sigma-level 
significance. Figure~\ref{fig:Ediff} gives the energy shifts of various processes
with associated statistical and systematic errors added in quadrature.

\section{Conclusion}

\noindent Lattice QCD calculations of two- and three-body interactions
of pions and kaons are now a precision science (for those channels
that do not involve disconnected diagrams). I have reviewed the
first fully-dynamical lattice QCD calculation of meson-baryon scattering.
An analysis of the scattering lengths of these two-body systems using
HB$\chi$PT has led to the conclusion that the three-flavor chiral
expansion does not converge over the range of light quark masses that
are investigated. By contrast, the $\pi^+\Sigma^+$ and
$\pi^+\Xi^0$ scattering lengths appear to have a well-controlled
chiral expansion in two-flavor HB$\chi$PT. Our results,
$a_{\pi^+\Sigma^+}=-0.197\pm0.017$ fm, and
$a_{\pi^+\Xi^0}=-0.098\pm0.017$ fm, deviate from the LO (current
algebra) predictions at the one- and two-sigma level, respectively. 

Clearly a milestone for this area of research would be to see a
definitive signal for nuclear physics. In the case of the NN
interaction, in addition to signal/noise issues, one must face a
fine-tuned system that requires a non-perturbative effective field
theory description. While this poses a tremendous challenge, this year
has seen a great deal of progress in the NPLQCD collaboration's state
goal to compute nuclear properties using lattice QCD.  During the last
year we performed a high-statistics calculation of one-, two- ---and
for the first time--- three-baryon
systems~\cite{Beane:2006mx,Beane:2006gf}, which represent a "jump" by
an order of magnitude in the data volume produced in such
calculations. The remarkable conclusion that the signal/noise problem
is not as severe as previously thought promises to hasten a golden age
of exploration for nuclear physics using lattice QCD.

\section*{Acknowledgments}

\noindent I thank W.~Detmold, T.C.~Luu, K.~Orginos, A.~Parre\~no,
M.J.~Savage, A.~Torok and A.~Walker-Loud for fruitful collaboration.
This work was supported by NSF CAREER Grant No. PHY-0645570.


\begin{thebibliography}{99}

\bibitem{Beane:2008dv}
  S.~R.~Beane, K.~Orginos and M.~J.~Savage,
  Int.\ J.\ Mod.\ Phys.\  E {\bf 17}, 1157 (2008)
  [arXiv:0805.4629 [hep-lat]].

\bibitem{Beane:2007xs}
  S.~R.~Beane, T.~C.~Luu, K.~Orginos, A.~Parreno, M.~J.~Savage, A.~Torok and A.~Walker-Loud,
  Phys.\ Rev.\  D {\bf 77}, 014505 (2008)
  [arXiv:0706.3026 [hep-lat]].

\bibitem{Bernard:2007zu}
  For a recent review, see V.~Bernard,
  Prog.\ Part.\ Nucl.\ Phys.\  {\bf 60}, 82 (2008)
  [arXiv:0706.0312 [hep-ph]].

\bibitem{Beane:2006mx}
  S.~R.~Beane, P.~F.~Bedaque, K.~Orginos and M.~J.~Savage,
  Phys.\ Rev.\ Lett.\  {\bf 97}, 012001 (2006)
  [arXiv:hep-lat/0602010].

\bibitem{Beane:2006gf}
  S.~R.~Beane, P.~F.~Bedaque, T.~C.~Luu, K.~Orginos, E.~Pallante, A.~Parreno and M.~J.~Savage
                  [NPLQCD Collaboration],
  Nucl.\ Phys.\  A {\bf 794}, 62 (2007)
  [arXiv:hep-lat/0612026].

\bibitem{Lepage:1989hd}
  G.~P.~Lepage, `The Analysis Of Algorithms For Lattice Field Theory,''
Invited lectures given at TASI'89 Summer School, Boulder, CO, Jun 4-30, 1989.
Published in Boulder ASI 1989:97-120 (QCD161:T45:1989).

\bibitem{Bedaque:2002mn}
  P.~F.~Bedaque and U.~van Kolck,
  Ann.\ Rev.\ Nucl.\ Part.\ Sci.\  {\bf 52}, 339 (2002)
  [arXiv:nucl-th/0203055].

\bibitem{Torok:2009dg}
  A.~Torok {\it et al.},
  arXiv:0907.1913 [hep-lat].

\bibitem{Beane:2009ky}
  S.~R.~Beane {\it et al.},
  Phys.\ Rev.\  D {\bf 79}, 114502 (2009)
  [arXiv:0903.2990 [hep-lat]].

\bibitem{Beane:2009gs}
  S.~R.~Beane {\it et al.},
  Phys.\ Rev.\  D {\bf 80}, 074501 (2009)
  [arXiv:0905.0466 [hep-lat]].

\bibitem{Bedaque:2007pe}
  P.~F.~Bedaque and A.~Walker-Loud,
  Phys.\ Lett.\  B {\bf 660}, 369 (2008)
  [arXiv:0708.0207 [hep-lat]].

\bibitem{Weinberg:1966kf}
  S.~Weinberg,
  Phys.\ Rev.\ Lett.\  {\bf 17}, 616 (1966).

\bibitem{Colangelo:2001df}
  G.~Colangelo, J.~Gasser and H.~Leutwyler,
  Nucl.\ Phys.\  B {\bf 603}, 125 (2001)
  [arXiv:hep-ph/0103088].

\bibitem{Leutwyler:2008fi}
  H.~Leutwyler,
  PoS C {\bf ONFINEMENT8}, 068 (2008)
  [arXiv:0812.4165 [hep-ph]].

\bibitem{Roy:1971tc}
  S.~M.~Roy,
  Phys.\ Lett.\  B {\bf 36}, 353 (1971).

\bibitem{Chen:2006wf}
  J.~W.~Chen, D.~O'Connell and A.~Walker-Loud,
  Phys.\ Rev.\  D {\bf 75}, 054501 (2007)
  [arXiv:hep-lat/0611003].

\bibitem{KaplanNelson}
  D.~B.~Kaplan and A.~E.~Nelson, preprint HUTP-86/A023;
%
  Phys.\ Lett.\  B {\bf 175} (1986) 57;
%
  Phys.\ Lett.\  B {\bf 192}, 193 (1987);
%
  Nucl.\ Phys.\  A {\bf 479}, 273 (1988);
%
  Nucl.\ Phys.\  A {\bf 479}, 285 (1988);

\bibitem{Lu:1994ex}
  M.~Lu, M.~B.~Wise and M.~J.~Savage,
  Phys.\ Lett.\  B {\bf 337}, 133 (1994)
  [arXiv:hep-ph/9407260].

\bibitem{Babich:2009rq}
  See, for instance, R.~Babich, R.~Brower, M.~Clark, G.~Fleming, J.~Osborn and C.~Rebbi,
  PoS {\bf LATTICE2008}, 160 (2008)
  [arXiv:0901.4569 [hep-lat]].

\bibitem{Jenkins:1990jv}
  E.~E.~Jenkins and A.~V.~Manohar,
  ``Baryon chiral perturbation theory using a heavy fermion Lagrangian,''
  Phys.\ Lett.\  B {\bf 255}, 558 (1991).

\bibitem{Liu:2006xja}
  Y.~R.~Liu and S.~L.~Zhu,
  Phys.\ Rev.\  D {\bf 75}, 034003 (2007)
  [arXiv:hep-ph/0607100].

\bibitem{Liu:2007ct}
  Y.~R.~Liu and S.~L.~Zhu,
  Eur.\ Phys.\ J.\  C {\bf 52}, 177 (2007)
  [arXiv:hep-ph/0702246].

\bibitem{Mai:2009ce}
  M.~Mai, P.~C.~Bruns, B.~Kubis and U.~G.~Mei\ss ner,
  arXiv:0905.2810 [hep-ph].

\end{thebibliography}
\end{document}